\documentstyle[epsfig]{mn2e}

\title[CC Sculptoris: A superhumping intermediate polar]{CC Sculptoris: A superhumping intermediate polar}
\author[P.A.~Woudt et al.]
       {P.A.~Woudt$^{1}$\thanks{email: Patrick.Woudt@uct.ac.za},
B. Warner$^{1,2}$, A. Gulbis$^{3,4}$, R. Coppejans$^{1,3}$, F.-J. Hambsch$^{5}$, \newauthor
A.P. Beardmore$^{6}$, P.A. Evans$^{6}$, J.P. Osborne$^{6}$, K.L. Page$^{6}$, G.A. Wynn$^{6}$, \& \newauthor
K. van der Heyden$^{1}$\\
        $^1$ Astrophysics, Cosmology and Gravity Centre,
        Department of Astronomy, University of Cape Town, Private Bag X3,\\
        Rondebosch 7701, South Africa\\
        $^2$ School of Physics and Astronomy, Southampton University, Highfield, 
        Southampton SO17 1BJ, UK\\
        $^3$ South African Astronomical Observatory, PO Box 9, Observatory 7935, South Africa\\
        $^4$ Southern African Large Telescope Foundation, PO Box 9, Observatory 7935, South Africa\\
        $^5$ Center for Backyard Astrophysics (Antwerp), American Association of Variable 
      Star Observers (AAVSO), \\ Vereniging Voor Sterrenkunde (VVS), Andromeda Observatory, Oude Bleken 12, 
      2400 Mol, Belgium \\
        $^6$ Department of Physics and Astronomy, University of Leicester, Leicester LE1 7RH, UK}
\date{Accepted 2012 August 27.  Received 2012 August 24; in original form 2012 July 30}

\begin{document}

\maketitle

\begin{abstract}
We present high speed optical, spectroscopic and \emph{Swift} X-ray observations made during 
the dwarf nova superoutburst of CC Scl in November 2011. An orbital period of 1.383 h 
and superhump period of 1.443 h were measured, but the principal new finding is that 
CC Scl is a previously unrecognised intermediate polar, with a white dwarf spin period 
of 389.49 s which is seen in both optical and \emph{Swift} X-ray light curves only during the 
outburst. In this it closely resembles the old nova GK Per, but unlike the latter
has one of the shortest orbital periods among intermediate polars.
\end{abstract}

\begin{keywords}
binaries: close -- novae, cataclysmic variables -- stars: oscillations -- stars: individual: CC Scl 
\end{keywords}

\section{Introduction}

CC Scl was listed as RX J2315.5-3049 in the ROSAT X-ray catalogue (Voges et al.~1999), 
later identified with a 17.3 magnitude star, classified as a cataclysmic variable (CV) 
(Schwope et al.~2000), and found in the Edinburgh-Cape Survey where it became EC 23128-3105 
(it is also in the 2MASS catalogue as 2MASS J23153185-3048476).  Its nature was confirmed 
when it was seen to rise in an outburst to magnitude 13.4 on 8 July 2000 (Stubbings 2000) 
followed by another about 100 days later. The latter was observed by Ishioka et al.~(2001) 
who suspected two apparent superhumps with a separation of 0.078 d. However, later photometric 
and spectroscopic observations made in quiescence (Augusteijn 2000; Chen et al.~2001; 
Tappert, Augusteijn \& Maza 2004; Augusteijn et al.~2010) gave an orbital period $P_{orb}$ of 0.0584 d 
(1.402 $\pm$0.005 h). Large amplitude flickering was noted and an occasional shallow narrow 
eclipse-like dip just after maximum light. Spectroscopic features included strong H and He\,\textsc{i} 
lines, with substantial strength of He\,\textsc{ii} 4686. Despite the last mentioned a classification 
simply as a dwarf nova was generally agreed upon.

   Ishioka et al.~(2001) noted that at 9 d duration the apparent superoutburst of 
CC Scl was unusually brief:  $\sim 14$ d is more to be expected of a short orbital period 
system. Following the two outbursts in 2000, CC Scl has had only seven outbursts reported by the 
AAVSO since (probably all rising to about the same maximum brightness), 
and only one outburst detected by the Catalina Real-Time Transient Survey 
(Drake et al.~2009) in the past 6 years -- the superoutburst of November 2011 discussed in 
this paper. Our observations were triggered by the CRTS announcement on 3 Nov 2011; 
PAW and BW were on the second night of a scheduled photometric observing run.

   In Sect.~2 we describe our optical observations of CC Scl and analyse these to 
find persistent periodicities. In Sect.~3 we show the Swift X-ray satellite observations 
that were obtained and in Sect.~4 we discuss the results and compare them with other CVs, 
in particular the very similar behaviour of Nova Persei 1901 (GK Per).

\section{Optical observations of CC Scl}

The data were taken using a new instrument called SHOC (Sutherland High-speed Optical Camera; 
Gulbis et al.~2011b).   The instruments were designed based on MORIS at NASA's 3-m Infrared 
Telescope Facility on Mauna Kea, Hawaii (Gulbis et al.~2011a). The primary components are (i) 
a high quantum efficiency ($>$ 90\% from roughly 480 nm to 700 nm), low read noise, low dark current, 
frame-transfer camera; (ii) a control computer capable of recording data at high rates, and (iii) 
a GPS for accurate timing.  SHOC uses an Andor iXon X3 888 UVB camera.  The UVB designation 
indicates a back-illuminated CCD with UV phosphor, elevating the quantum efficiency to approximately 
35\% below 380 nm, and an uncoated fused silica window. The camera has a selection of readout amplifiers 
including conventional and electron-multiplying (EM) modes, each having multiple preamplifer gain settings.

For these observations, SHOC was mounted below the filter wheel on the 40-inch (1.0-m) reflector at 
the Sutherland site of the South African Astronomical Observatory (SAAO). 
The field of view was 2.9 arcmin $\times$ 2.9 arcmin, and the data 
were binned 2x2 for a plate scale of 0.34$''$/pix.  The 1 MHz EM mode was employed with an EM 
gain of 20 and a preamplifier gain of 7.5 $e^-$/ADU. These settings have an effective read noise 
on the order of 1 $e^-$/pix.  The camera was thermoelectrically cooled at $-70$ C, resulting in 
a dark current of less than 0.001 $e^-$/pix/sec.

\begin{table}
 \centering
  \caption{Observing log of photometric observations obtained with the SHOC camera at 
the South African Astronomical Observatory, South Africa.}
   \begin{tabular}{@{}llccrc@{}}
 Run      & Date of obs.          & HJD first obs. & Length  & t$_{in}$  &  V \\
          &                       &  (+2450000.0)     & (h)     & (s)      & (mag) \\[10pt]
S8119 & 2011 Nov 03 & 5869.24749 & 2.80 &  1 & 13.4   \\
S8121 & 2011 Nov 04 & 5870.23977 & 2.75 &  1 & 13.6   \\
S8123 & 2011 Nov 05 & 5871.23970 & 3.34 &  4 & 13.8   \\
S8124 & 2011 Nov 06 & 5872.24102 & 5.05 &  4 & 14.2   \\
S8126 & 2011 Nov 07 & 5873.24092 & 4.56 &  4 & 14.2   \\
S8128 & 2011 Nov 08 & 5874.25021 & 4.52 &  4 & 14.1   \\[5pt]
\end{tabular}
\label{woudttab1}
\end{table}

\begin{table}
 \centering
  \caption{Observing log of photometric observations obtained at the Remote Observatory Atacama
Desert (ROAD) in Chile.}
   \begin{tabular}{@{}llccrc@{}}
 Run      & Date of obs.          & HJD first obs. & Length  & t$_{in}$  &  V \\
          &                       &  (+2450000.0)     & (h)     & (s)      & (mag) \\[10pt]
H01   & 2011 Nov 04 & 5870.49069 & 6.27 & 30 & 13.6   \\
H02   & 2011 Nov 05 & 5871.48644 & 6.38 & 45 & 13.9   \\
H03   & 2011 Nov 06 & 5872.48561 & 6.39 & 45 & 14.3   \\
H04   & 2011 Nov 07 & 5873.48613 & 6.38 & 45 & 14.2   \\
H05   & 2011 Nov 08 & 5874.48708 & 6.34 & 45 & 14.2   \\
H06   & 2011 Nov 09 & 5875.51221 & 5.73 & 45 & 15.6   \\
H07   & 2011 Nov 10 & 5876.49337 & 6.13 & 90 & 16.3   \\
H08   & 2011 Nov 11 & 5877.49508 & 6.08 & 90 & 16.5   \\
H09   & 2011 Nov 12 & 5878.52106 & 5.39 & 90 & 16.6   \\
H10   & 2011 Nov 13 & 5879.49426 & 5.81 &120 & 16.6   \\
H11   & 2011 Nov 14 & 5880.49137 & 5.75 &120 & 16.8   \\
H12   & 2011 Nov 15 & 5881.49045 & 5.75 &120 & 16.6   \\
H13   & 2011 Nov 16 & 5882.49037 & 5.74 &120 & 16.8   \\[5pt]
\end{tabular}
\label{woudttab2}
\end{table}

All the measurements were made without any optical filter, 
have been calibrated by using known UV-rich standard stars and corrected for 
atmospheric extinction in the usual manner. A log of the South African (SA) observational 
runs is given in Tab.~\ref{woudttab1}; we were fortunate in having six consecutive clear nights, 
which provided excellent coverage of CC Scl.

In addition we have an extensive set of light curves obtained by FJH 
at the Remote Observatory Atacama Desert (ROAD) in Chile, using a 
commercially available CCD camera from Finger Lakes Instrumentation, 
an FLI ML16803 CCD. The Microline (ML) line of CCD cameras is a lightweight 
design which can hold a variety of CCD chips. The full frame Kodak 
KAF-16803 image sensor used together with the 40 cm Optimized Dall Kirkham (ODK) f/6.8 
telescope from Orion Optics UK gives a field of view of nearly 48 arcmin x 48 arcmin, 
and the data were binned 3x3 for a plate scale of $2.09''$/pix to keep 
the amount of data recorded to a reasonable value. The download speed was 8 MHz, the 
preamplifier gain 1.4 $e^-$/ADU. The camera was thermo-electrically cooled to a 
temperature of $-25$ C. All the measurements are made with a photometric V filter. 
The log of these observations 
is given in Tab.~\ref{woudttab2}.

\begin{figure}
\centerline{\hbox{\psfig{figure=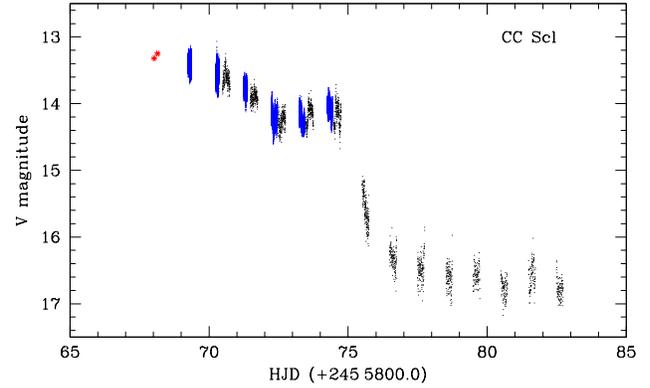,width=8.4cm}}}
  \caption{The long term light curve of CC Scl. The red asterisks 
mark the CRTS observations, the blue dots correspond to the SAAO data
and the black dots are observations obtained in Chile (colour only available
in the online version).}
 \label{lcccscllt}
\end{figure}

  The outburst light curve of CC Scl, combined from the CRTS, SAAO and Chile 
observations, is shown in Fig.~\ref{lcccscllt}. The early `plateau' part of the curve 
lasted at least 7 days (HJD 2455868 -- 75); there is no record of when the outburst 
commenced but we suggest in Sect.~2.2 that it probably occured on 2 November (HJD 2455868). 
The rapid decay phase lasts only about 2 days, but there is a slow reduction after 
that (probably due to the post-outburst cooling of the white dwarf) to the CRTS 
long-term quiescent magnitude of 16.8.  These are compatible with the total of 9 
days recorded by Ishioka et al.~(2001), and provides a range of outburst of 3.6 mag, 
so the duration and amplitude are unusually small for a short orbital period 
superoutbursting CV.

\begin{figure}
\centerline{\hbox{\psfig{figure=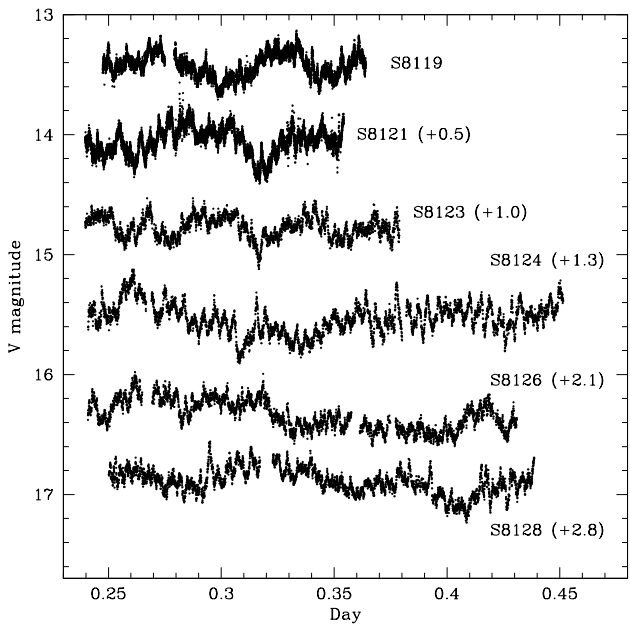,width=8.4cm}}}
  \caption{Individual light curves of CC Scl taken with the SHOC camera on consecutive
nights during the outburst plateau (3 -- 8 November 2011). The light 
curve of run S8119 is displayed at the correct brightness; vertical offsets for each
light curve are given in brackets.}
 \label{lcccsclshoc}
\end{figure}

\begin{figure}
\centerline{\hbox{\psfig{figure=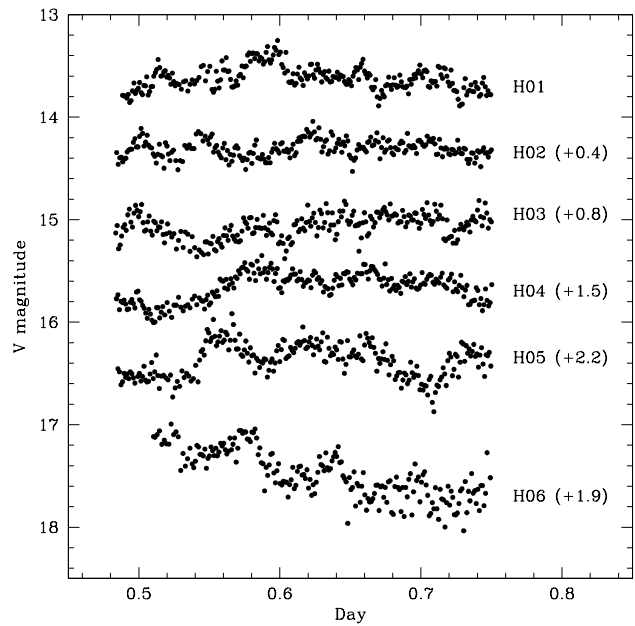,width=8.4cm}}}
  \caption{Individual light curves of CC Scl taken at the Remote Observatory
Atacama Desert in Chile on consecutive nights during outburst and decline (4 -- 9 November 2011). The light 
curve of run H01 is displayed at the correct brightness; vertical offsets for each
light curve are given in brackets.}
 \label{lcccsclhambsch1}
\end{figure}

  The full set of SAAO light curves, all taken during the outburst, is shown 
in Fig.~\ref{lcccsclshoc} and those obtained in Chile during outburst in Fig.~\ref{lcccsclhambsch1}.
It is obvious in the suite of light curves that there are both slow variations on time scales 
$\sim 100$ min in the earliest runs, and a very prominent $\sim 400$ s 
modulation later on the plateau. We first examine the evidence for orbital and 
superoutburst periods in our entire data set.

\subsection{Modulations at orbital time scales.}

\begin{figure}
\centerline{\hbox{\psfig{figure=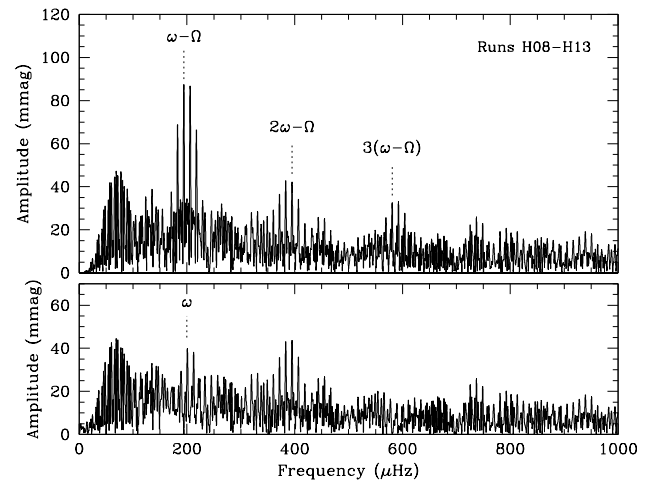,width=8.4cm}}}
  \caption{The Fourier transform of six combined nights at quiescence (runs H08--H13 in Tab.~\ref{woudttab2}).
 $\omega$ is the orbital frequency, $\Omega$ is the precession frequency
of the disc. The lower panel is of the light curve prewhitened at frequencies $\omega - \Omega$ and
3($\omega - \Omega$).}
 \label{fthambsch}
\end{figure}

To search for any modulation at the known orbital period we look at the 
quiescent ROAD photometry (Chile), starting with 11 November (runs H08-H13, Tab.~\ref{woudttab2}). 
The Fourier transform (FT) of these 6 nights is shown in Fig.~\ref{fthambsch}; the 
dominant frequency is a superhump and its harmonics, which we analyse below. 
To detect the orbital modulation we prewhiten at the superhump fundamental and
its second harmonic (see below), 
which reveals a probable very weak orbital modulation, seen in the lower 
panel of Fig.~\ref{fthambsch}. The indicated period is 1.383 ($\pm$ 0.001) h 
at an amplitude of 40 mmag, very close to the Chen et al.~(2001)
spectroscopic period which was based on very limited observations.
The amplitude of the variations in the light curve during outburst (Fig.~\ref{lcccsclshoc}, and Sect.~2.2)
will obscure any shallow eclipse as occasionally seen in earlier observations (Chen et al.~2001).
The individual light curves in quiescence are too noisy to see such a shallow eclipse.

\begin{figure}
\centerline{\hbox{\psfig{figure=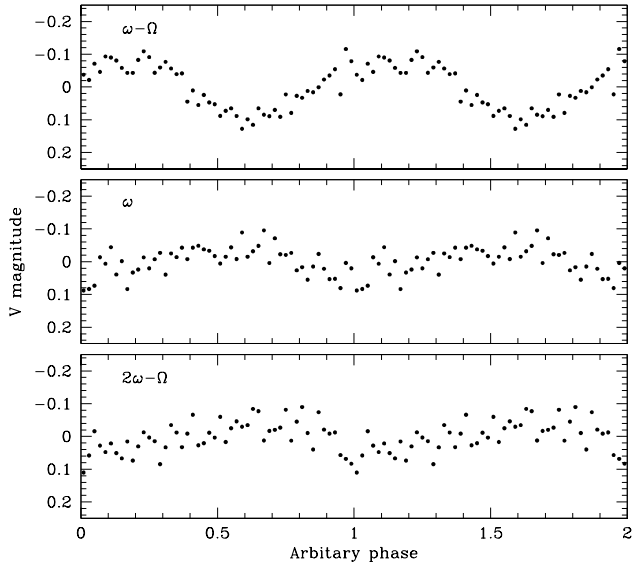,width=8.4cm}}}
  \caption{Upper panel: The average light curve of the late superhump in quiescence, marked
by $\omega - \Omega$ in Fig.~\ref{fthambsch}. Middle panel: The average light curve at the
orbital frequency ($\omega$) after the data have been prewhitened at $\omega - \Omega$ and 
$3 (\omega - \Omega$). Bottom panel: The average light curve at the orbital sideband
of the superhump frequency, $2 \omega - \Omega$, after the light curve has been prewhitened
at $\omega - \Omega$, $3 (\omega - \Omega)$ and $\omega$.
All panels are based on the combined runs H08--H13 during the return to quiescence.}
 \label{ccscllcav}
\end{figure}

   Strong superhumps are present throughout the outburst and quiescent 
light curves. In studies of superhumps the notation used for components in the amplitude spectra of 
light curves is usually $\omega$ for the orbital frequency and $\Omega$ for 
the precession frequency of the elliptical disc. The superhump frequency is 
then $\omega - \Omega$, and harmonics, together with sum and difference 
frequencies, are commonly seen. Our FT for the runs during superoutburst (runs H01-H05 in 
Tab.~\ref{woudttab2}, see also the lower panel of Fig.~\ref{ftallcomb}) 
shows a strong signal at a period of 1.443 h, which, combined with the 
orbital period of 1.383 h gives a precession period for the disc of 1.4 d. 
During quiescence for the week immediately after superoutburst we again find a 
dominant signal at the superhump period, now 1.430 h (Fig.~\ref{fthambsch}), 
implying that the precession period has lengthened to 1.8 d. 
But in addition to $\omega - \Omega$ there are now peaks at $\omega$, 
$2 \omega - \Omega$  and $3(\omega - \Omega)$. In well-studied 
superoutbursts frequencies such as $3\omega - \Omega$ and higher orders 
are often detected (e.g. IY UMa: Patterson et al.~2000a; DV UMa: Patterson 
et al.~2000b), but $2\omega - \Omega$ is not detectable. However, it does appear 
in some (helium-transferring) AM CVn stars (e.g. HP Lib: Patterson et al.~2002; 
AM CVn: Skillman et al.~1999). We have no explanation of why this orbital 
sideband of the superhump frequency appears only in CC Scl among H-rich CVs.
The average profiles of the three signals seen in Fig.~\ref{fthambsch}
are shown in Fig.~\ref{ccscllcav}.

The 6 nights after outburst (runs H08-H13, Tab.~\ref{woudttab2}) show an alteration of mean brightness
(Fig.~\ref{lcccscllt}). A sine curve fit to these nights produces a period of 1.96 d and
amplitude of 0.17 mag. This is generated by beats between the 1.430 h 88 mmag superhump
and 1.383 h 40 mmag orbital modulation, i.e.~the $\Omega$ [= $\omega - (\omega - \Omega)$]
frequency.

\begin{figure}
\centerline{\hbox{\psfig{figure=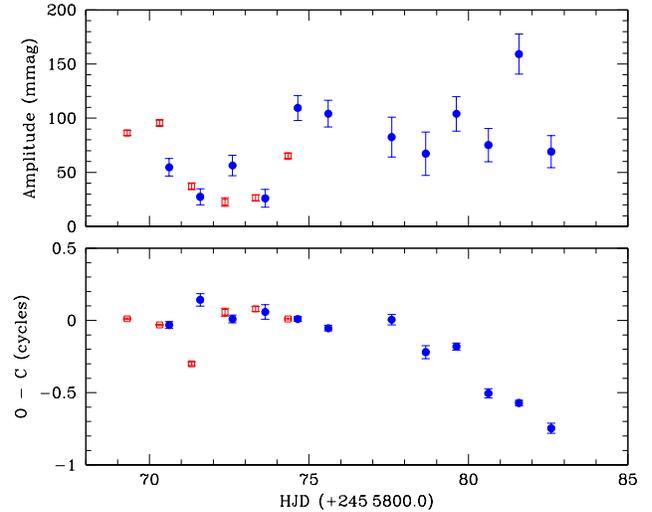,width=8.4cm}}}
  \caption{The amplitude and phase variations of the 1.443-h superhump modulation
as determined from the Chile data (filled circles) and the SAAO observations (open squares). The
zero point for the phase is at HJD 2455870.5000; this corresponds to the minimum 
light of the superhump modulation.}
 \label{omcccsclsh}
\end{figure}

In superoutbursting dwarf novae (SU UMa stars) it has previously 
been reported that superhumps persist for several days after outburst, as ``late superhumps'', 
with  period similar to that during outburst, but shifted in phase by $\sim 180^{\circ}$ 
(e.g. DV UMa: Patterson et al.~2000a). But this may be quite rare because most ``late superhumps'' 
appear after a switch to a shorter period (or a steadily changing phase) without 
any phase discontinuity (e.g. SDSS J0137: Pretorius et al.~2004; ‘Stage C’ in Kato et al.~2009; 
V344 Lyr: Wood et al. 2011).
This behaviour is seen in CC Scl (Fig.~\ref{omcccsclsh}). We have omitted the run H07 from this 
plot as no definite superhump was detectable.

\begin{figure}
\centerline{\hbox{\psfig{figure=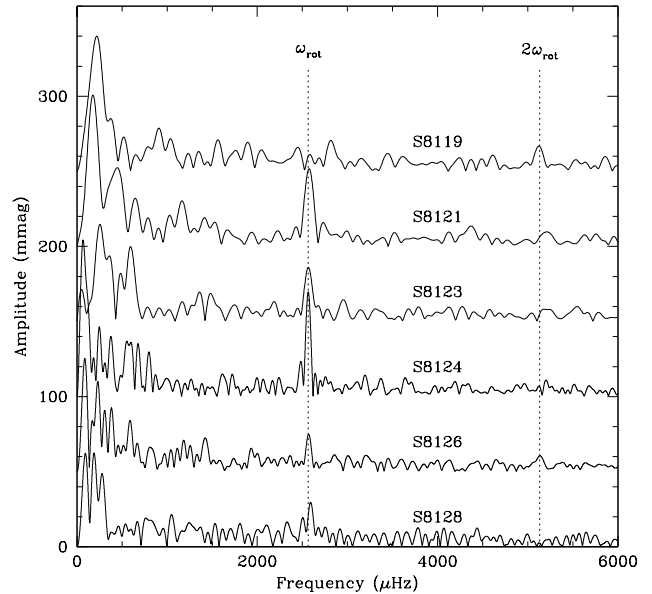,width=8.4cm}}}
  \caption{The Fourier transform of six individual runs (S8119 - S8128) obtained with SHOC
during superoutburst.}
 \label{ftsss2315ind}
\end{figure}

\subsection{Modulations at the 390 s time scale}

\begin{figure}
\centerline{\hbox{\psfig{figure=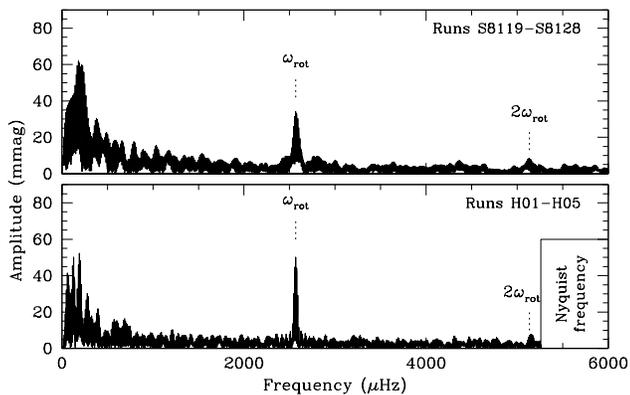,width=8.4cm}}}
  \caption{Top: The Fourier transform of six individual runs (S8119 - S8128) obtained with SHOC
during superoutburst. Bottom: Fourier transform of five combined runs (H01 - H05) obtained with
ROAD over the same interval.}
 \label{ftallcomb}
\end{figure}

Fourier transforms of all of the outburst light curves, with the exception of the 
first (on 3 November 2011), show a prominent peak at 389.5 s previously unreported from this object. 
On 3 Nov the first harmonic of this fundamental period is quite strongly present. The FTs of the 
Sutherland individual runs are shown in Fig.~\ref{ftsss2315ind}. An FT 
for all of them combined and one for the suite of Chilean outburst runs 
is given in Fig.~\ref{ftallcomb}. 
It is interesting to see how the latter observations, acquired with a 40 cm aperture 
and despite the modulation being invisible to the eye in the light curves (Fig.~\ref{lcccsclhambsch1}), 
show the 390 s signal with great prominence - a valuable consequence of the lengths 
of the runs being twice those of the Sutherland ones.

   None of the FTs show any orbital sidebands to the fundamental period, and 
as the optical and X-ray periods (see Sect.~3) are identical we conclude that 
the modulation arises from the rotation period ($P_{rot} = 1/\omega_{rot}$) of a magnetic white dwarf 
primary, which puts CC Scl into the intermediate polar (IP) class of CVs 
(see Chapter 7 of Warner (1995)). An explanation for the absence of the fundamental 
period on the first night of our observations is then available by analogy with the 
model of the outburst properties of the IP XY Ari (Hellier, Mukai \& Beardmore 1997): 
on the first night the outburst had started only about a day before and the wave 
of increased mass transfer through the disc had not yet reached the inner regions, leaving 
two accretion zones optically visible, but by the second night the radius of the 
magnetosphere that was keeping the inner parts of the disc open had been reduced 
by increased gas flow, obscuring the lower accretion zone on the white dwarf. 
The shallow eclipses seen in quiescence (Sect.~1) show that CC Scl is of the 
moderately high inclination required for this explanation. Note, however, that there are 
occasional weak signals at the first harmonic during the outburst (Figs.~\ref{ftsss2315ind}
and \ref{ftallcomb}), but 
this could arise from a slightly non-sinusoidal profile to the fundamental signal, 
rather than direct viewing of the second accretion pole.

From the combined sets of observations we find the following ephemeris for
maximum light of the 390-s modulation

$${\mathrm{HJD_{max}}} = 245\,5870.241331 + 0.00450801 (6) \, E. $$

\subsection{Optical spectroscopy during superoutburst}

During run S8124 (see Tab.~\ref{woudttab1}) simultaneous spectroscopy with the 
Cassegrain spectrograph on the 74-in reflector of the SAAO was obtained. Over
a two hour period (HJD 2455872.279 - 2455872.364), eleven 600-s spectra were recorded. 
For these observations grating 7 was used with a slit width of $1.8''$, giving a 
wavelength coverage of 3800 -- 7738 {\AA} at 2.28 {\AA} per pixel.
Relative flux corrections were made using the spectrophotometric
standard LTT\,9293 and absolute calibrations were performed based on the simultaneous
observations obtained with the SHOC camera (see Tab.~\ref{woudttab1}).

\begin{figure}
\centerline{\hbox{\psfig{figure=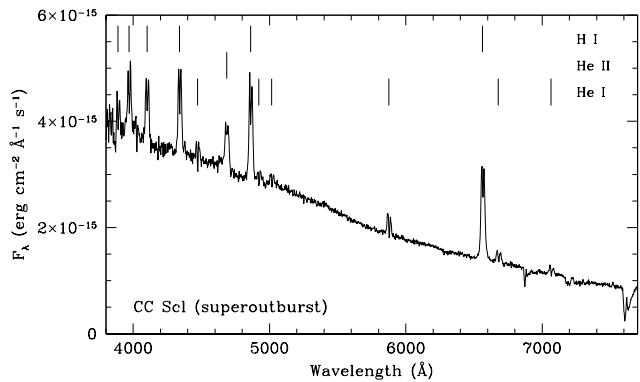,width=8.4cm}}}
  \caption{The averaged optical spectrum of CC Scl obtained during superoutburst. The 
identified emission lines are marked by the vertical lines.}
 \label{ccsclspectrum}
\end{figure}

The combined spectrum is shown in Fig.~\ref{ccsclspectrum} and shows, besides the double-peaked
Balmer emission lines, strong He\,\textsc{ii} 4686 {\AA} emission as well as numerous double-peaked 
He\,\textsc{i} lines. All the identified lines are indicated by the vertical markers in Fig.~\ref{ccsclspectrum}.
Compared with the spectrum of CC Scl in quiescence (Chen et al. 2001; Tappert et al. 2004), 
the He\,\textsc{ii} line has been greatly enhanced, becoming about half instead of $\sim$10\% the 
intensity of H$\beta$, and the 4650 {\AA} C\,\textsc{iii}/N\,\textsc{iii} blend which is not definitely 
detectable in quiescence, is clearly present on the blue wing of 4686 {\AA}. Although 
more commonly associated with magnetic CVs, these enhancements can occur in 
superoutbursts of ordinary dwarf novae (e.g. Z Cha: Honey et al. 1988). All the lines 
are doubled, which indicates a large inclination and is compatible with the presence 
of the shallow eclipses in quiescence.

\begin{figure}
\centerline{\hbox{\psfig{figure=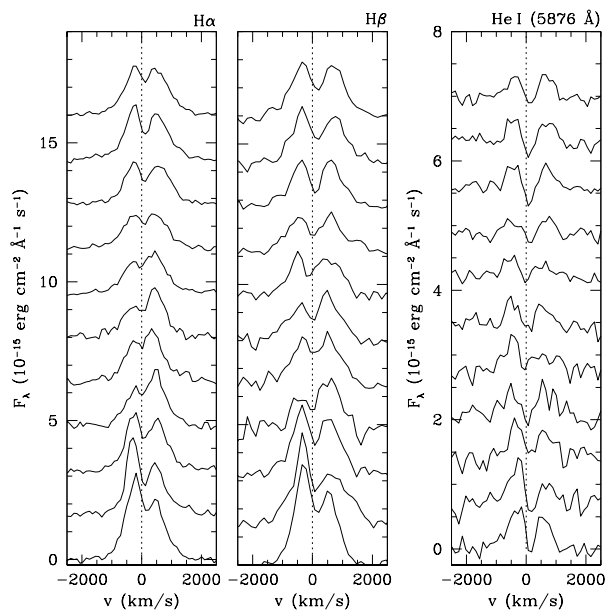,width=8.0cm}}}
  \caption{Variability observed in the individual lines across the orbital cycle. From 
bottom to top, consecutive individual spectra are shown centred on lines of H$\alpha$, 
H$\beta$ and He\,\textsc{i} 5876 {\AA}. Each spectrum covers $\sim$ 0.12 in orbital phase.
Incremental offsets of $1.6 \times 10^{-15}$ and 
$0.7 \times 10^{-15}$ erg cm$^{-2}$ {\AA}$^{-1}$ s$^{-1}$ have been applied to the Balmer lines 
and He\,\textsc{i} line respectively, for display purposes.}
 \label{ccsclspectravar}
\end{figure}

   In early studies of IPs an important clue to their structure was given by the 
orbital modulation of emission line strengths and their components -- e.g. the V/R 
ratio of the doubled lines. The V/R ratio is a relative measurement of the 
equivalent widths on either side of the rest wavelength of the line. Through modelling 
the velocity-amplitude of the observed S-wave in the emission lines as originating
from the disc-stream impact region, it proved the existence of an accretion disc 
impacted on its rim by an accretion stream, as in non-magnetic CVs (e.g. FO Aqr: 
Hellier, Mason \& Cropper 1990; AO Psc: Hellier, Cropper \& Mason 1991). Our spectra 
show these same characteristics, seen in H$\alpha$, H$\beta$ and He\,II (Fig.~\ref{ccsclspectravar}), 
but our time resolution is insufficient to allow folding on spin phase, which was 
a direct way of demonstrating rotation of the accreting magnetospheres in the FO Aqr 
and AO Psc analyses.

\section{Swift observations}

Although the \emph{Swift} satellite (Gehrels et al. 2004) was designed for 
the study of gamma-ray bursts, its rapid-response capabilities also make 
it ideal for following other transient sources such as novae and cataclysmic 
variables. There are three instruments onboard \emph{Swift}: the wide-field BAT 
(Burst Alert Telescope; Barthelmy et al. 2005), which covers 15 -- 350 keV; 
and the narrow-field instruments: the XRT (X-ray Telescope; Burrows et al. 2005) 
spanning 0.3 -- 10 keV; and the UVOT (UV/Optical Telescope; Roming et al. 2005) 
with filters covering 1700 -- 6500 {\AA}. Swift Target of Opportunity 
observations were requested shortly after CC Scl was found to be in a 
superoutburst, these were continued until it ended. Here we present 
results from the XRT and UVOT, CC Scl was not bright enough to have been 
detected by the BAT. \emph{Swift} has an orbital period of 1.6 h, source 
visibility constraints prevent continuous observations longer than $\sim$0.5 h; 
for these reasons \emph{Swift} has reduced sensitivity to the orbit and 
superhump periods of CC Scl. 

\emph{Swift} observed CC Scl between JD 2455873 -- 81, starting 5 days after 
outburst if this occurred on Nov 2; Tab.~\ref{obslogswift} lists the observations 
obtained. All XRT data were in photon counting mode (which provides two-dimensional 
imaging, and is typically used for sources with count rates below $1-2$ count s$^{-1}$, 
while the UVOT was operated in event mode and the data collected using the uvw2 filter 
(central wavelength of 2246 {\AA}; FWHM of 498 {\AA}). Version 3.8 of the \emph{Swift}
software, corresponding to HEASoft 6.11 (released 2011-06-07), was used, together 
with the most recent version of the Calibration Database. In our XRT analysis we 
used only grade 0 events to avoid pile-up.

\begin{table}
 \centering
  \caption{Log of the \emph{Swift} observations in 2011.}
   \begin{tabular}{@{}lcccc@{}}
Seg.  & Start time & End time & Time since & Exp. \\
No.   & (UTC)    & (UTC)      & first obs. (ks) & time (ks) \\[10pt]
001   & 07 Nov at 15:57 & 16:22 & 0      & 1.48 \\
002   & 08 Nov at 11:11 & 16:23 & 69.2   & 5.45 \\
003   & 09 Nov at 00:12 & 00:36 & 116.1  & 1.40 \\
004   & 09 Nov at 02:03 & 02:10 & 122.8  & 0.38 \\
005   & 09 Nov at 03:09 & 03:33 & 126.7  & 1.43 \\
006   & 09 Nov at 04:56 & 05:22 & 133.2  & 1.54 \\
007   & 10 Nov at 00:03 & 00:26 & 202.0  & 1.27 \\
008   & 10 Nov at 01:38 & 02:02 & 207.7  & 1.44 \\
009   & 10 Nov at 03:18 & 03:43 & 213.7  & 1.50 \\
010   & 10 Nov at 04:50 & 05:15 & 219.2  & 1.48 \\
011   & 11 Nov at 00:12 & 02:13 & 289.7  & 3.04 \\
012   & 11 Nov at 13:06 & 14:58 & 336.1  & 2.96 \\
013   & 12 Nov at 00:11 & 02:11 & 375.3  & 2.31 \\
014   & 12 Nov at 13:07 & 15:10 & 422.6  & 3.10 \\
015   & 13 Nov at 00:16 & 02:17 & 462.7  & 2.98 \\
016   & 13 Nov at 13:10 & 15:12 & 509.0  & 2.74 \\
017   & 14 Nov at 00:21 & 02:22 & 549.4  & 2.88 \\
018   & 14 Nov at 13:12 & 15:16 & 595.6  & 2.86 \\
019   & 15 Nov at 00:27 & 02:27 & 636.1  & 2.74 \\
020   & 15 Nov at 11:42 & 13:44 & 676.6  & 2.73 \\
\end{tabular}
\label{obslogswift}
\end{table}

\begin{figure}
\centerline{\hbox{\psfig{figure=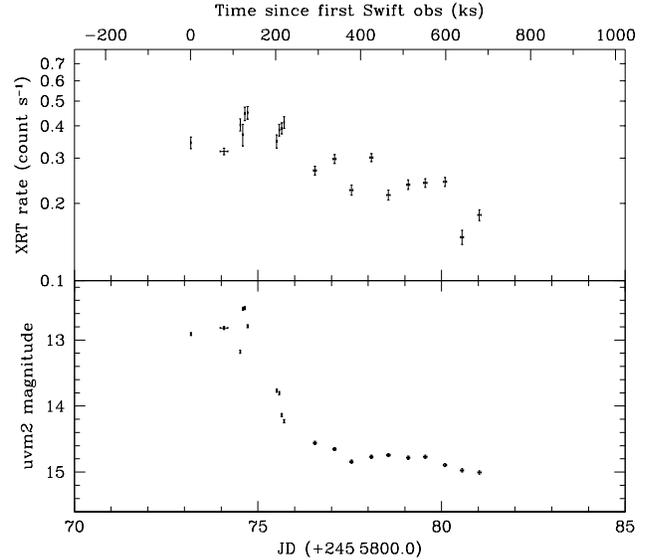,width=8.4cm}}}
  \caption{Comparison of the XRT (0.3--10 keV; upper panel) and the
UVOT (uvm2 filter; lower panel) light curves, showing that the evolution
is correlated, see also Fig.~\ref{lcccscllt}. The \emph{Swift}
data are plotted with one bin per Obs ID.}
 \label{lcccscluvx}
\end{figure}

\subsection{X-ray spectral analysis}

\begin{figure}
\centerline{\hbox{\psfig{figure=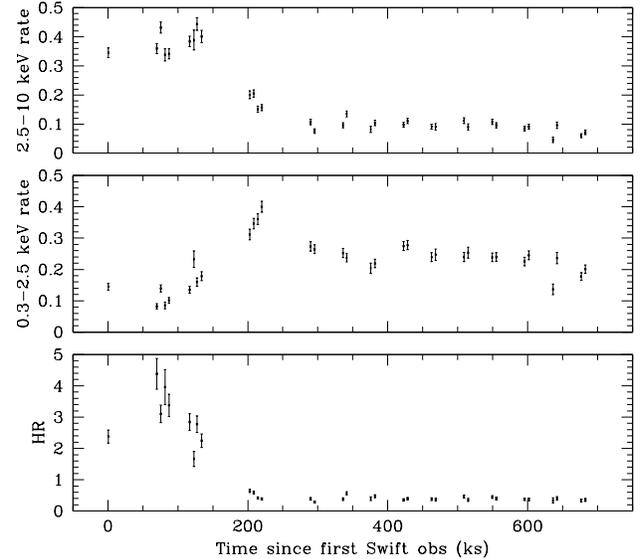,width=8.4cm}}}
  \caption{XRT hardness ratio (lower panel), comparing the 0.3 -- 2.5 keV (middle panel)
and 2.5 -- 10 keV (upper panel) bands, with one bin per snapshot. The emission clearly 
softens about 200 ks after the first observation.}
 \label{ccsclhr}
\end{figure}

Fig.~\ref{lcccscluvx} shows the \emph{Swift} X-ray and UV light curves, these can 
be compared with the optical light curve in Fig.~\ref{lcccscllt}. The X-rays 
show a general declining trend, while the UV evolution more closely matches 
that in the optical, declining sharply around JD 2455875. The X-ray hardness 
ratio ($HR = 2.5-10$ keV/$0.3-2.5$ keV) is shown in Fig.~\ref{ccsclhr}, there 
is a marked spectral softening around the time of the sharp optical-UV decline.

\begin{figure}
\centerline{\hbox{\psfig{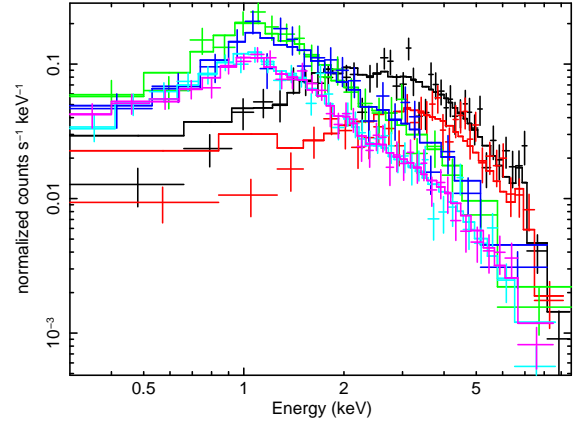}}}
  \caption{\emph{Swift} XRT spectra extracted for the phase of X-ray spin maximum and minimum for
segments 001--006 (max = black, min = red), 007--010 (max = blue, min = green) and
011--020 (max = cyan, min = magenta). Fits are given in Tab.~\ref{coverswift}. Colour
available only in the online version.}
 \label{ccsclxrayminmax}
\end{figure}

Time-resolved X-ray spectra were created, both for each observation (at late 
times, observations were merged to improve the signal-to-noise), and covering the 
minimum and maximum 0.2 cycles of the spin modulation (Section 2.2) before and 
after the sharp optical-UV decline. These can all be well fitted with a solar abundance, 
partially covered cooling flow model (see Tabs.~\ref{coolswift} and \ref{coverswift}, 
and Fig.~\ref{ccsclxrayminmax}). Simpler models do not fit or are physically inappropriate. 
The cooling flow model ({mkcflow} in \textsc{xspec}) has a distribution of 
emission measure against 
temperature which is the inverse of the bolometric luminosity for optically thin plasma 
emission, and is commonly used to represent the shocked accretion column as it cools and 
settles onto a magnetic white dwarf. In this model the lower and upper temperatures were 
fixed at 0.1 and 30 keV respectively; no interstellar absorption was needed.

\begin{table}
 \centering
  \caption{Partially-covered cooling flow model ({mkcflow} in {\textsc{xspec}}), 
with the low and high temperature limits set to 0.1 and 30 keV respectively.
Segment number as in Tab.~\ref{obslogswift}. At later times, consecutive segments
have been combined to improve the signal-to-noise.}
   \begin{tabular}{@{}cccc@{}}
Seg.  & $N_H$                   & Covering & C-stat/dof\\
No.   & ($10^{22}$ cm$^{-2}$)    & fraction &           \\[10pt]
001   & 2.77$^{+0.46}_{-0.40}$  &  0.97 $\pm$ 0.01         & 308/393\\
002   & 4.37$^{+0.35}_{-0.35}$  &  $0.985^{+0.003}_{-0.004}$ & 659/674 \\
003   & 3.16$^{+0.58}_{-0.48}$  &  0.98 $\pm$ 0.01         & 335/397 \\
004   & 1.86$^{+0.45}_{-0.38}$  &  $>$ 0.97               & 157/167 \\
005   & 3.40$^{+0.53}_{-0.45}$  &  0.98 $\pm$ 0.01        & 267/381 \\
006   & 2.80$^{+0.48}_{-0.41}$  &  0.98 $\pm$ 0.01        & 276/342 \\
007   & 0.60$^{+0.20}_{-0.15}$  &  $0.91^{+0.05}_{-0.06}$   & 285/320 \\
008   & 0.71$^{+0.31}_{-0.22}$  &  0.79 $\pm$ 0.06        & 341/377 \\
009   & 0.43$^{+0.30}_{-0.21}$  &  $0.66^{+0.15}_{-0.11}$   & 292/333 \\
010   & 0.49$^{+0.31}_{-0.20}$  &  0.60 $\pm$ 0.10        & 272/358 \\
011   & 0.24$^{+0.25}_{-0.14}$  &  $0.59^{+0.27}_{-0.16}$   & 369/381 \\
012-013 & 0.61$^{+0.33}_{-0.22}$ &  0.61 $\pm$ 0.07       & 463/479 \\
014-015 & 0.14$^{+0.25}_{-0.07}$ & $>$ 0.48               & 457/482 \\
016-017 & 0.65$^{+0.33}_{-0.24}$ &  0.53 $\pm$ 0.07       & 441/467 \\
018-020 & 0.57$^{+0.62}_{-0.34}$ &  $0.38^{+0.11}_{-0.08}$  & 395/489 \\
\end{tabular}
\label{coolswift}
\end{table}

\begin{table}
 \centering
  \caption{As Tab.~\ref{coolswift}, but for \emph{Swift} XRT spectra extracted for 
the minimum and maximum of the spin period over time.}
   \begin{tabular}{@{}ccccc@{}}
Seg.  & min/ & $N_H$                   & Covering & C-stat/dof\\
No.   & max    & ($10^{22}$ cm$^{-2}$)    & fraction &           \\[10pt]
001-006   &  max     & 3.94$^{+0.41}_{-0.33}$  &  $0.966^{+0.005}_{-0.006}$ & 395/420 \\
001-006   &  min     & 7.64$^{+0.94}_{-0.73}$  &  0.981 $\pm$ 0.004       & 259/314 \\
007-010   &  max     & 0.25$^{+0.15}_{-0.11}$  & $>$ 0.65                 & 194/214 \\
007-010   &  min     & 0.52$^{+0.21}_{-0.16}$  &  0.77 $\pm$ 0.08         & 181/217 \\
011-020   &  max     & 0.10$^{+0.12}_{-0.01}$  & $>$ 0.66                 & 350/363 \\
011-020   &  min     & 0.44$^{+0.18}_{-0.13}$  &  0.55 $\pm$ 0.07         & 332/375 \\
\end{tabular}
\label{coverswift}
\end{table}

\begin{figure}
\centerline{\hbox{\psfig{figure=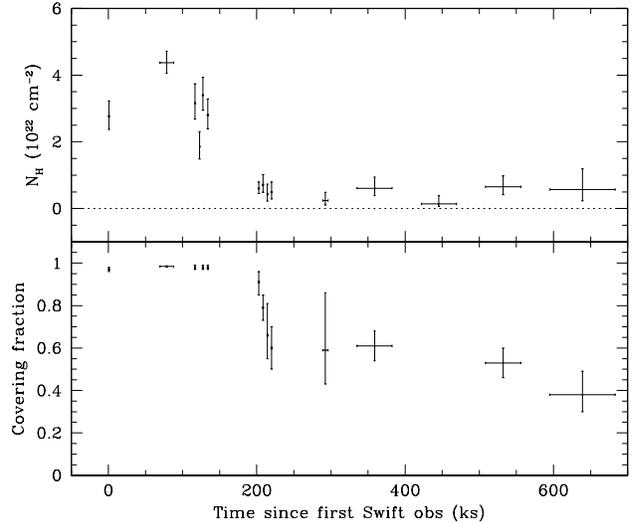,width=8.4cm}}}
  \caption{The variation with time of absorbing column (upper panel) and covering
fraction (lower panel), derived from \emph{Swift} XRT spectral fits
as given in Tab.~\ref{coolswift}. The dashed horizontal line in the 
upper panel indicates $N_H = 0$.}
 \label{ccsclabscol}
\end{figure}

\begin{figure}
\centerline{\hbox{\psfig{figure=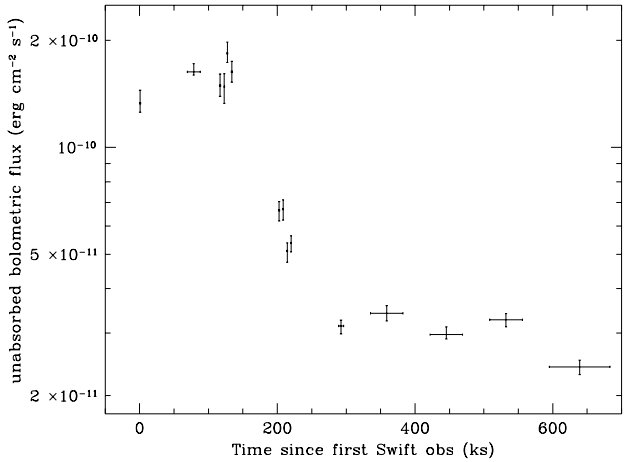,width=8.4cm}}}
  \caption{The unabsorbed bolometric (0.01 -- 200 keV) flux calculated from the \emph{Swift} XRT spectral
fits as a function of time. The error bars were estimated by using the limits on the
$N_H$ uncertainties.}
 \label{ccsclunabs}
\end{figure}

The good fits we have achieved demonstrate that the intrinsic emission of 
CC Scl is not significantly changing its spectral distribution during the observations. 
However, both the absorption column and covering fraction show a clear decline 
coincident with the optical-UV flux decline (see Fig.\ref{ccsclabscol}). This 
change in absorption has a substantial effect on the fraction of the intrinsic 
flux that we see. In Fig.~\ref{ccsclunabs} we show the bolometric flux ($0.01 - 200$ keV) 
light curve of CC Scl created by removing the effect of this changing absorption, a 
decline in luminosity by a factor $\sim 5$ at the time of the optical-UV decline 
is clearly evident. Tab.~\ref{coverswift} and Fig.~\ref{ccsclxrayminmax} show 
that CC Scl is always more absorbed at the time of the observed flux minimum in 
the spin cycle, but that this effect is much stronger when the outburst was at its brightest.

\subsection{389.5 s modulation}

\begin{figure*}
\centerline{\hbox{\psfig{figure=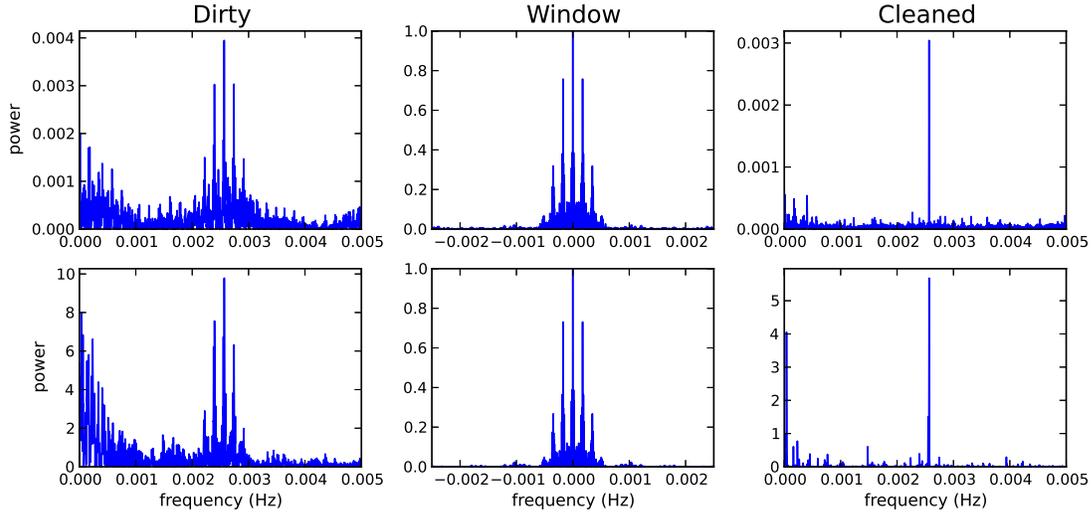,width=16cm}}}
  \caption{Upper panels: Cleaned X-ray power spectrum covering the Obs IDs 001--006.
Lower panels: Cleaned UV power spectrum, up to Obs IDs 010.}
 \label{ccsclxuvft}
\end{figure*}

\begin{figure}
\centerline{\hbox{\psfig{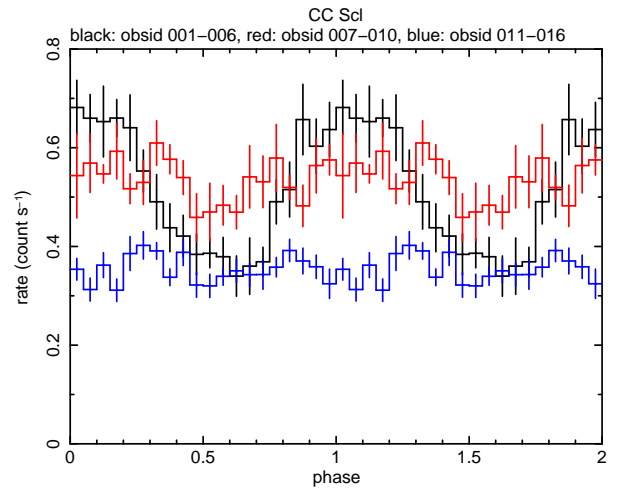}}}
  \caption{The \emph{Swift} XRT folded light curves in three epochs (black, red and blue
being progressively later in time; colours in online version only). The data
are folded on the optical emphemeris given in Sect.~2.2. }
 \label{ccsclxrayfold}
\end{figure}

\begin{figure}
\centerline{\hbox{\psfig{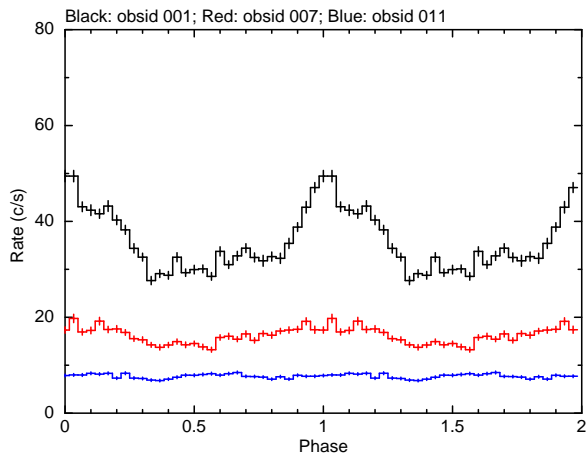}}}
  \caption{The UVOT data at three fiducial epochs, folded on the optical ephemeris given
in Sect.~2.2.}
 \label{ccscluvepoch}
\end{figure}

We used the CLEAN procedure (Robert, Lehar \& Dreher 1987; Norton, Beardmore \& Taylor 1996)
to generate power spectra of the observed X-ray and UV count rates binned at 50 and 5 s
respectively and de-trended with a second order polynomial. The upper panels of 
Fig.~\ref{ccsclxuvft} show the results for the first six X-ray observations; the lower 
panels plot the UVOT results for data up to observation segment 010. There is a 
very clear detection of the 389.5 s modulations during these time intervals. The 
period is measured from the X-ray clean power spectrum to be $389.5 \pm 0.4$ s, while 
the UV value is $389.3 \pm 0.4$ s. Both values are consistent with the period measured 
more precisely in the optical data. The evolution of the X-ray pulse is illustrated by 
Fig.~\ref{ccsclxrayfold}, which shows how the modulation fraction declined before 
the observed count rate. Fig.~\ref{ccscluvepoch} shows three epochs of UVOT data folded 
on the optical ephemeris, revealing an initial fractional amplitude similar to that 
seen in X-rays, although this declines along with the flux eventually falling to 
a much greater extent than is seen in the X-ray count rate (the UV decline is much 
more consistent with the bolometric X-ray flux decline seen in Fig.~\ref{ccsclunabs} 
however). As is also the case in X-rays, the 389.5 s modulation is not detected at 
late times in the UV.

\begin{figure}
\centerline{\hbox{\psfig{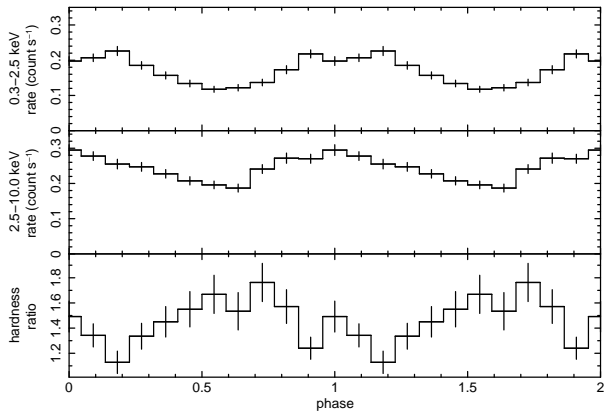}}}
  \caption{The \emph{Swift} XRT folded hardness ratio (see also Fig.~\ref{ccsclhr} for an $HR$ plot),
showing the classic IP behaviour of an anti-correlation between intensity and $HR$.
The $HR$ covers 2.5 -- 10 keV/0.3 -- 2.5 keV and includes the segments up to and
including 010.}
 \label{ccsclhr2}
\end{figure}

Consistent with the phase resolved spectral fits, the plot of the hardness ratio 
as a function of 389.5 s phase in Fig.~\ref{ccsclhr2} clearly demonstrates the 
characteristic IP behavior:  the source is harder when fainter. Both the X-ray 
and UV modulation profiles peak at the time of the optical maximum (like AO Psc, 
but unlike PQ Gem; Evans \& Hellier 2005a), suggesting a common origin in all bands, 
and thus that the accretion curtains are obscuring the hot accretion column across 
the entire spectrum at this time. 

We know that a viscous accretion disk must be present in CC Scl because we 
see superhumps, it therefore follows that the 389.5 s period must be the spin 
period of the magnetic white dwarf. While it is possible for X-rays to show 
an orbital beat of the spin period rather than the spin period itself, this is 
very rare and can only happen if the white dwarf accretes from a structure fixed 
in orbital phase. Although V2400 Oph can show this behavior (Buckley et al. 1995), 
this system is supposed not to have a viscous disk (Hellier \& Beardmore 2002); 
thus we conclude that in CC Scl $P_{spin} = 389.5$ s.

There is a strong decline in the spin-modulated flux fraction in both X-rays 
and UV as the luminosity of CC Scl decreases from outburst to quiescence. As 
discussed in Sect. 2.2, these changes can be due to a higher accretion rate during 
the outburst forcing the inner disc radius in towards the white dwarf, blocking 
the lower accretion pole so that we can see the light curve modulated at the white 
dwarf spin period. Then, when the accretion rate decreases again, the inner disc 
radius moves back out, meaning that we can see the lower pole, the emission from 
which is in anti-phase with the upper pole and effectively cancels the modulation 
(e.g. Hellier et al.~1997). The view to the lower pole could also be blocked in 
outburst if the disk were to thicken at this time.

\section{Discussion}

\subsection{CC Scl as an Intermediate Polar}

The list of definitely confirmed IPs (Mukai 2011) contains thirty-five members, 
to which we can now add CC Scl. Only five of those in Mukai's catalogue have 
$P_{orb} < 2$ h, i.e. are below the `orbital period gap'; CC Scl is thus 
an addition to this rare group. There are, however, two additional very short
$P_{orb}$ systems that should be given full IP status: V455 And (Matsui et al. 2009; 
Silvestri et al. 2012) and WZ Sge (Warner \& Pretorius 2008).
In Tab.~\ref{tableips} we list their properties, 
clustered into three groups: the slow rotators with $P_{orb}/P_{spin} \sim 2$, intermediate
with $P_{orb}$/$P_{spin} \sim 10$ and those with $P_{orb}$/$P_{spin} \sim 100$. 

CC Scl is only the second example of the middle group (see Mukai (2011) for a complete plot of 
$P_{orb}$ versus $P_{spin}$ in which the rarity of low $P_{orb}$ systems is clearly visible). 
The first two groups represent respectively spins that are near equilibrium with angular momentum 
transferred directly from near to the inner Lagrangian point by strong magnetic fields, 
or, for weaker fields, after passage through an accretion disc which acquires angular 
momentum more appropriate to the circularisation radius (King \& Wynn 1999). We deduce 
from this that the white dwarfs in HT Cam and CC Scl have magnetic moments significantly 
lower than those in the first four systems, and that V455 And and WZ Sge are lower still.

\begin{table}
 \centering
  \caption{Intermediate Polars below the `orbital period gap'.}
   \begin{tabular}{@{}lcccc@{}}
Star & P$_{orb}$ (h) & $P_{spin}$ (s) & $P_{orb}$/$P_{spin}$ & $\Delta T$ (d) \\[10pt]
SDSS J2333+15   &      1.385   &         2500   &      1.99  &         -   \\
V1025 Cen         &      1.41    &         2147   &      2.36  &      discless \\
DW Cnc            &      1.44    &         2315   &      2.24  &        $2-4$  \\   
EX Hya            &      1.64    &         4022   &      1.47  &        $2-3$  \\[5pt]
CC Scl            &      1.38    &          389   &     12.8   &             9 \\
HT Cam            &      1.43    &          515   &     10.0   &             3 \\[5pt]    
V455 And         &  1.35        & 67.6         &  72   &    17 \\
WZ Sge           &  1.36        & 27.8         & 176   &    25 \\
\end{tabular}
\label{tableips}
\end{table}

  Tab.~\ref{tableips} also notes dwarf nova-like outburst durations $\Delta T$ of the group of short 
periods IPs (none reported in SDSS\,J2333+15). V1025 Cen is believed to have a sufficiently strong field that no 
standard disc forms - instead mass transfer takes place through a centrifugally supported 
torus which does not suffer the standard disc instability. DW Cnc, EX Hya, and HT Cam have 
very brief infrequent outbursts, also unlike standard dwarf novae, which probably result 
from intermittent storage of gas just outside the magnetically defined inner edge of the 
accretion disc (Spruit \& Taam 1993); therefore CC Scl is the only member of the middle group which 
has almost canonical disc instabilities, even though of slightly short durations for superoubursts.
V455 And and WZ Sge have only superoutbursts.

The magnetosphere of the white dwarf in CC Scl will produce an inner disc radius of 
$r_0 \sim 1.3 \times 10^{10}$ cm, which is $\sim 15 R(1)$ and $\sim 0.6 r_d$, where the notation 
and equations 2.61, 2.83a, 7.17 of Warner (1995) have been used. Thus about 40\% of the outer 
disc of CC Scl is available for regular dwarf nova outbursts (but only $\sim 25$\% of the 
outer disc of HT Cam). This may account for (a) the shorter than normal superoutbursts 
in CC Scl (there is less mass to drain out) and (b) the extremely short outbursts in HT Cam. 
These formulae do not apply to the longest $P_{spin}$ group in Tab.~\ref{tableips} because 
the structure of what amounts to an accretion torus rather than a truncated disc is not 
known - but the effect evidently is to allow only very short-lived outbursts, and this may be 
what is happening in HT Cam as well.

From analysing X-ray spectra, Evans \& Hellier (2005b) find that EX Hya, V1025 Cen 
and HT Cam are the only IPs that have low surface density accretion curtains 
in quiescence (with $N_H < 2 \times 10^{21}$ cm$^{-2}$), implying a low rate of mass 
transfer in these systems, as is expected for CVs below the period gap. For CC Scl 
our X-ray spectra show $N_H > 4 \times 10^{21}$ cm$^{-2}$ at spin minimum in outburst 
and quiescence, perhaps suggesting a higher accretion rate even though it too 
has an orbital period below the gap. Taking the quiescent luminosity of CC Scl from 
the latter part of Fig.~\ref{ccsclunabs}, we have $L \sim 3.6 \times 10^{33} D^2$ erg s$^{-1}$, 
where $D$ is the distance in kpc. Considering the accretion rate to be $RL/GM$, and 
assuming a white dwarf mass of 0.6 M$_{\odot}$, we find the accretion rate to 
be $\sim 7 \times 10^{-10} D^2$ M$_{\odot}$ yr$^{-1}$. The high Galactic latitude of 
CC Scl ($b = -69^\circ$) suggests  $D \ll 1$, but without a better measure we 
cannot decide whether the accretion rate in CC Scl is high for its orbital period 
(for example EX Hya, with $P_{orb}$ close to that of CC Scl, has an accretion rate 
of $6 \times 10^{-11}$ M$_{\odot}$ yr$^{-1}$, Beuermann et al. 2003), and thus cannot 
test the absorption -- accretion rate relation suggested by Evans \& Hellier.

\subsection{Comparison with GK Per}

GK Per was Nova Persei 1901, one of the brightest novae of the twentieth century. 
It fell to its pre-eruption brightness after about 11 years, but since 1966 it has 
shown dwarf nova outbursts with an amplitude of 2 to 3 mag roughly every 3 years. 
In addition, it is established as an IP from a 351.3 s modulation in hard X-rays 
during outbursts (Watson, King \& Osborne 1985), which is of low amplitude or absent 
during quiescence. However, it is seen at low amplitude in the photometric U band, 
even during quiescence, but is less coherent at longer wavelengths and better described 
as quasi-periodic near 380 s (Patterson 1991).  During outbursts the GK Per spectral 
lines stay in emission, rather than changing to broad absorption lines as in normal dwarf novae. 

   The similarities to CC Scl are marked, but one way in which GK Per differs 
totally from CC Scl is that it has an orbital period that is one of longest known 
for normal CVs: it is 1.997 d, deduced from spectroscopy.  This implies a large 
separation of components and a very large disc radius, of which only a small part 
of the central region is swept clear by the rotating magnetosphere of the primary. 
An effect of the large $P_{orb}$ is that it is difficult to make photometric observations 
that show any orbital brightness modulation, and even more difficult to see superhumps, 
if they exist. The possibility of superhumps arises because the secondary in GK Per is 
a subgiant reduced in mass through evolution and consequent mass transfer. The estimated 
mass ratio $q$ in GK Per is $0.55 \pm 0.21$ (Morales-Rueda et al.~2002), so using the formulae 
$r_d/a = 0.6/(1+q)$ and $r_3 = 0.46a$ for the quiescent outer disc radius and the 3-to-1 
resonance radii, respectively (Warner 1995), we have $r_3 \sim 1.19 r_d$ for $q = 0.55$, and 
therefore the outer radius of the disc has only to expand by $\sim$20\% (to take up the 
angular momentum released by infalling matter) during a dwarf nova outburst for the 
disc to start evolving into an elliptical shape. The precession period $P_{prec}$ of the 
disc for $q = 0.55$ is $\sim (10 - 12) P_{orb}$, i.e. $\sim 20 - 24$ d in GK Per. Although there 
have been no obvious superoutbursts among the dwarf nova outbursts in GK Per, 
the 2006 outburst (Evans et al.~2009) was noted as unusually long ($\sim 100$ d), and we note that there were three maxima 
spaced about 30 d apart, which might be evidence for an effect of $P_{prec}$. At $\sim 60$ d 
the duration of most of the GK Per outbursts is too short for a $\sim 30$-d periodicity 
to be properly established.

   The X-ray, spectral and photometric similarities between GK Per and CC Scl 
suggest that the latter might have had a classical nova eruption sometime in the 
past few hundred years. There is no detectable nova shell in the availableç direct 
images, and GALEX images do not show any resolvable shell. An HST snapshot to look 
for H$\alpha$ emission nebulosity could be justified.

\subsection{Superoutbursts in intermediate polars}

Superoutbursts among IPs have been considered rare, perhaps even non-existent, 
e.g., Patterson et al.~(2011) have recently assessed the literature and conclude that 
although several IPs have dwarf nova outbursts, ``no confirmed superhumpers are known 
to be magnetic''. Nevertheless there are a couple of examples that might qualify: 
Patterson et al.~(2011) themselves cite TV Col and RX0704 as possible cases, but WZ Sge, 
the most extreme of the SU UMa stars, with outstanding superoutbursts and superhumps, 
has rapid oscillations ($\sim 28$ s) that appear in optical and X-rays and which have been 
interpreted with an IP model (e.g. Warner, Livio \& Tout 1996; Warner \& Pretorius 2008). 
To this we can now certainly add CC Scl, as an unambiguous example of a superhumping IP, 
and from our earlier discussion GK Per could be a candidate. It is not clear from the 
long-term light curve of CC Scl (which, apart from the 2011 outburst, is very poorly 
sampled) whether all the outbursts except the latest are ordinary ones or whether 
they are superoutbursts. With two outbursts only 100 d apart it would be unusual to 
find they are both superoutbursts, but there is no sign in the light curve of a clear 
division in outburst ranges, and the fact that the only one (and of the usual maximum 
brightness), monitored with sufficient time resolution, turns out to have superhumps 
suggests that they all could be superoutbursts.

  The question raised by CC Scl is why it alone, among the short orbital period IPs 
that have dwarf nova outbursts, is able to generate an elliptical disc and associated 
superhumps. The answer presumably lies in whatever limits outbursts in the other systems 
to only a couple of days in length - as a result there probably isn't sufficient 
time to get the eccentric mode fully excited.

\section*{Acknowledgments}
PAW and BW acknowledge support from the University of Cape Town (UCT) and from the
National Research Foundation (NRF) of South Africa. This work was completed during a
visit by PAW to the University of Southampton, which was funded by an ERC advanced
investigator grant awarded to R Fender. 
APB, PAE, JPO and KLP acknowledge financial support from the UK Space Agency.
AG acknowledges support from the NRF.
We thank Andrea Dieball for looking 
in the GALEX archive at the position of CC Scl.
We thank the \emph{Swift} PI and team for their support of these
Target of Opportunity observations.
This paper uses observations made at the South African Astronomical Observatory.

\end{document}